\documentclass[]{aa}
\usepackage[dvips]{graphicx}
\usepackage{times}
\usepackage{natbib}
\def\ebv{\mbox{$E_{B-V}$\,}}
\def\halpha{\mbox{H$\alpha$}}
\def\hbeta{\mbox{H$\beta$}}
\def\hgamma{\mbox{H$\gamma$}}
\def\peryr{\mbox{$\>\rm yr^{-1}$}}
\def\subsun{\mbox{$_{\normalsize\odot}$}}
\def\lesssim{\mathrel{\hbox{\rlap{\hbox{\lower4pt\hbox{$\sim$}}}\hbox{$<$}}}}

\begin{document}

\title{The Host Galaxy of \object{GRB 990712}\thanks{Based on observations
    collected at the European Southern Observatory, La Silla, with the ESO
    3.6m telescope, NTT, and VLT (ESO Programmes 165.H-0464(I) and
    165.H-0464(E), and 265.D-5742(B)) by the Gamma-Ray Burst Afterglow
    Collaboration at ESO (GRACE) group. Further based on observations with the
    Danish 1.5m telescope.}$^,$\thanks{Based on observations with the NASA/ESA
    Hubble Space Telescope, obtained at the Space Telescope Science Institute,
    which is operated by the Association of Universities for Research in
    Astronomy, Inc. under NASA contract NAS5-26555.}}

\author{Lise~Christensen\inst{1}
  \and Jens~Hjorth\inst{2}
  \and Javier~Gorosabel\inst{3,4,5} 
  \and Paul~Vreeswijk\inst{6,7}
  \and Andrew~Fruchter\inst{8}
  \and Kailash~Sahu \inst{8}
  \and Larry~Petro \inst{8}
}
\institute{Astrophysikalisches Institut Potsdam, An der Sternwarte 16, 14482 
 Potsdam, Germany
  \and Niels Bohr Institute, Astronomical Observatory, University of
  Copenhagen, Juliane Maries Vej 30, 2100 Copenhagen {\O}, Denmark
  \and Instituto de Astrof\'{\i}sica de Andaluc\'{\i}a, IAA-CSIC, 
 Granada, Spain
  \and Laboratorio de Astrof\'{\i}sica Espacial y F\'{\i}sica Fundamental 
 (LAEFF-INTA), P.O. Box 50727, 28080, Madrid, Spain 
  \and Danish Space Research Institute, Juliane Maries Vej 30, 
 2100 Copenhagen {\O}, Denmark
  \and European Southern Observatory, Casilla 19, Santiago, Chile
  \and Astronomical Institute "Anton Pannekoek'', University of Amsterdam 
 \& Center for High Energy Astrophysics, Kruislaan 403, 1098 SJ Amsterdam, 
 The Netherlands
  \and Space Telescope Science Institute, 3700 San Martin Drive, 
 Baltimore, MD 21218.
}
          
\mail{lchristensen@aip.de}
\date{Received / Accepted}

\abstract{ We present a comprehensive study of the $z=0.43$ host galaxy of
  \object{GRB 990712}, involving ground-based photometry, spectroscopy, and
  HST imaging. The broad-band $U\!BV\!RI\!J\!H\!K\!s$ photometry is used to
  determine the global spectral energy distribution (SED) of the host galaxy.
  Comparison with that of known galaxy types shows that the host is similar to
  a moderately reddened starburst galaxy with a young stellar population.  The
  estimated internal extinction in the host is $A_V=0.15\pm 0.1$ and the
  star-formation rate (SFR) from the UV continuum is
  1.3$\pm$0.3~M\subsun\peryr\, (not corrected for the effects of extinction).
  Other galaxy template spectra than starbursts failed to reproduce the
  observed SED. We also present VLT spectra leading to the detection of
  \halpha\, from the GRB host galaxy. A SFR of 2.8$\pm$0.7 M\subsun\peryr\, is
  inferred from the \halpha\, line flux, and the presence of a young stellar
  population is supported by a large equivalent width. Images from HST/STIS
  show that the host has two separate knots, which could be two distinct
  star-forming regions.
\keywords{gamma rays: bursts -- galaxies: hosts -- galaxies: star formation}}

\maketitle

\section{Introduction}
For all but one of the Gamma-Ray Bursts (GRBs) where the position of the
X-ray, optical, or radio afterglow has been localised to an accuracy of less
than 1\arcsec, follow-up deep observations have revealed underlying galaxies.
The current sample consists of $\sim$45 such GRB hosts, and $\sim$35 of these
have measured redshifts in the range $0.168<z<4.5$ \citep{hjorth03,andersen00}
and have magnitudes 21~$<~R~<$~30. The faintness of the hosts requires long
integration times on the largest telescopes to obtain high signal to noise
ratio spectra. Ground-based broad-band photometry presents a useful
alternative for investigating the spectral energy distribution (SED).

The relatively small size of GRB hosts sometimes makes it difficult to tell
which morphological type it is, e.g. whether the radial intensity profile of
the galaxy is best fitted by an exponential disk profile or an elliptical
profile. However, studies of some hosts with the HST have shown that an
exponential profile provides a good fit to the surface intensity distribution
\citep{ode98,fruchter00,hjorth02}. Comparing the SED of the host with the SEDs
of known galaxy types provides an alternate method of estimating the galaxy
type. \citet{sok01} analysed 6 GRB hosts in this way, showing that all of them
had SEDs characteristic of starburst galaxies. This is expected if GRBs are
associated with massive collapsing stars as suggested by \citet{woosley93} and
\citet{wijers98} and recently observed for the \object{GRB~030329}
\citep{stanek03,hjorth03}. A small age for the burst population gives an
indication that GRB progenitors are massive stars, whereas SED ages much
longer than the life times of the most massive stars could indicate a binary
merging event as the cause of the GRB \citep{eichler89}.

The GRB 990712 host is bright relative to other GRB hosts, and therefore
serves as a good case for studying the multi-band SED. This paper is one in a
series of papers on the SEDs of GRB hosts. Studies of the hosts of
\object{GRB~000210} and \object{GRB~000418} are presented in \citet{goro02}
and \citet{goro03}, respectively.
  
The previous studies of the \object{GRB~990712} afterglow and the host are
summarised in Sect.~\ref{burst}. In Sect.~\ref{data} and Sect.~\ref{halpha} we
present photometry and spectroscopy of the \object{GRB~990712} host. The
morphology of the host is investigated in Sect.~\ref{morph}. A comparison of
the SED derived from all the observations with spectral synthetic templates is
described in Sect.~\ref{sed}. In Sect.~\ref{SFR} we estimate the SFR of the
host galaxy using two SFR estimators; first using the UV continuum and second
the \halpha\, line flux.  In Sect.~\ref{disc} we discuss the results.

Throughout the paper we assume $\Omega_m=0.3$, $\Omega_{\Lambda}=0.7$ and
$H_0=65$~km~s$^{-1}$~Mpc$^{-1}$. At the redshift of the host, $z=0.433$, the
luminosity distance is $d_L=7.93\times 10^{27}$cm.

\section{\object{GRB 990712}}
\label{burst}

\object{GRB 990712} was detected on July 12 1999 at UT 16:43:02, by the
Italian-Dutch satellite BeppoSAX \citep{gcn385}.  The burst had the strongest
afterglow observed in X-rays to that date. Its afterglow was found
approximately 4 hours after the trigger by \citet{gcn387}. A spectrum obtained
shortly after revealed a redshift of $z=0.433$ from the emission lines
[\ion{O}{II}], [\ion{O}{III}], H$\gamma$ and H$\beta$ as well as \ion{Mg}{I}
and \ion{Mg}{II} absorption lines \citep{gcn388,hjorth00}. The relatively low
redshift makes it one of the closest GRB hosts. Only the \object{GRB~980425}
at $z=0.0085$ \citep{gal98}, \object{GRB~011121} at $z=0.36$ \citep{infante01}
and \object{GRB~030329} at $z=0.168$ \citep{greiner03} were nearer.

A spectrum of the combined flux from the host and the afterglow was obtained
by \citet{vrees01} 1.5 days after the burst. Because of the brightness of the
host, the spectrum shows distinct absorption lines and emission lines from the
host itself.

The [\ion{O}{II}] emission line flux was measured to be
(3.37$\pm$0.2)~$\times~10^{-16}$~erg~cm$^{-2}$~s$^{-1}$ \citep{vrees01}, which
corresponds to a SFR of 2.7$\pm0.8$~M$_{\odot}$yr$^{-1}$ using the conversion
from measured flux to a SFR from \citet{kennicutt98} (Hereafter K98).
Converting the flux at restframe 2800~{\AA} to a SFR gives a similar result.
An internal extinction of $A_V~=~3.4^{+2.4}_{-1.7}$ was inferred from the flux
ratio of the hydrogen lines \hgamma/\hbeta. The host was observed at radio
frequencies (1.4 GHz) by \citet{vrees01b}, who did not find any radiation from
the host to a limit of 70 $\mu$Jy. This upper limit implies that the total
unextincted SFR in the host is less than 100~M$_{\odot}$~yr$^{-1}$. This is in
great contrast to the measurements of the host of \object{GRB~980703}, which
was found to have a SFR of $\sim$500~M$_{\odot}$yr$^{-1}$ measured from its
radio flux \citep{berger01,berger02}.  The \object{GRB 990712} host has an
IR-luminosity which is 20 times less than that of the luminous
\object{GRB~980703} host. This suggests that different types of galaxies can
host GRBs some having more dust enshrouded star formation than others.

\section{Imaging and photometry}
\label{data}
We have analysed both ground based and HST images of the host of \object{GRB
  990712}. The ground-based observations consist of $U\!BV\!RI\!J\!H\!K\!s$
images obtained at different dates and using different instruments. All of the
data presented here were obtained more than one year after the burst, so that
the flux contribution from the afterglow is negligible. The data obtained from
the Danish 1.5-m in September 2000 consist of images in Bessel $B, V$ and $R$
and Gunn $I$ filters, and the $U$ band data were obtained at the ESO 3.6-m
telescope the night of Aug. 13, 2001 using the EFOSC2 instrument. Near-IR
$J\!H\!K\!s$ images were obtained at the NTT with the SOFI instrument over two
nights from Aug.  1--2, 2001. Only the first night was photometric according
to the ESO webpage\footnote{\tt http://www.eso.org/gen-fac/pubs/astclim/
  forecast/meteo/CIRA/images/repository/lossam/}.

In order to obtain a more reliable optical estimate of the SFR of the host of
\object{GRB 990712}, we performed spectroscopic observations centered on
H$\alpha$, using FORS2 on the VLT.

The following sections describe the details of the data reduction,
calibrations and combination of the different data sets.

\subsection{HST/STIS images}
The HST/STIS images of the host of \object{GRB 990712} were obtained on April
24, 2000 as a part of a survey of GRB hosts \citep{fruchter00b}\footnote{The
  FITS files of this and other hosts can be downloaded at {\tt
    http://www.ifa.au.dk/$\sim$hst/grb\_hosts/intro.html}}. The GRB hosts are
observed through a clear (unfiltered) aperture (the 50CCD filter, which in the
following analyses will be called the CL filter) and a long pass imaging
filter, F28$\times$50LP (called the LP filter).

The sensitivity of the CL filter extends from 4000~{\AA} to 9000~{\AA}, with
its peak at $\sim$5800~{\AA}, which falls within the $V$ passband.  The
sensitivity of the LP filter extends from 5500~{\AA} to 9000~{\AA}, with its
peak at $\sim$6000~{\AA}. The total integration time of the
\object{GRB~990712} host was 4080 s in each filters.

The individual images are combined using DITHER II, a package which includes
several tasks needed for combining dithered HST images.  \citet{fruchter02}
describe the drizzling of WFPC2 images, and we adopt this process for the
HST/STIS images. This method allows a higher resolution in the final images
than in the original STIS images. In the drizzling of the STIS images of the
hosts, the parameters {\tt pixfrac}=0.6 and {\tt scale}=0.5 were used. This
gives an output pixel size of 0\farcs0254 pixel$^{-1}$ in a 2k$\times2$k
frame.

\subsection{Ground-based optical images}
The DFOSC data in the $B,V, R$ and $I $ filters consist of 11, 15, 15, and 59
frames having total integration times of 12900~s, 10800~s, 7600~s, and
39950~s, respectively. The individual integration times in each filter were
not of equal length.

The raw images were bias subtracted after inspection of the overscan region of
the CCD. Any residual structure in the bias level was corrected for using a
normalised median filtered bias image. The images were flat-fielded using flat
fields obtained from a combination of several twilight sky observations. The
selected flat fields produced reduced images in which the background signal
varied by less than 2\%, which is the limit for DFOSC data. The reduced images
were WCS calibrated with a pipeline written by Andreas Jaunsen (ESO,
Santiago). Approximately 50 reference stars from the USNO2 catalogue were used
to compute the astrometry. From the WCS calibrated images, the shifts and
rotations between individual images were found. The images were drizzled in
much the same way as the HST images, not altering the pixel scale so the pixel
size of the output image is 0\farcs39 pixel$^{-1}$.

The drizzling method was tested by comparing the final drizzled image with a
median filtered image. The drizzling method gave a higher S/N for faint
objects compared to the median combined image.

Another problem to take into account is the fringes that appear in the $I$
band images and which depend on background level and consequently on the
integration time. Although the fringes are large scale structures in the DFOSC
images and should not introduce a large error when combining 59 individual
images, they can be removed by construction of a fringe frame image. Such a
frame is produced by replacing the pixel values of all stars with a small
pixel value, while also replacing their neighboring pixels and combining the
images using a threshold rejection. The resulting fringe image was scaled to
the exposure time of each image and subtracted. This removed the fringes from
the images in almost all cases. In a few ($\sim$5) images the background was
higher, and the fringes were removed by subtracting a scaled fringe frame,
with a scale factor higher than the integration times.

\begin{figure}
\centering
\resizebox{\hsize}{!}{\includegraphics{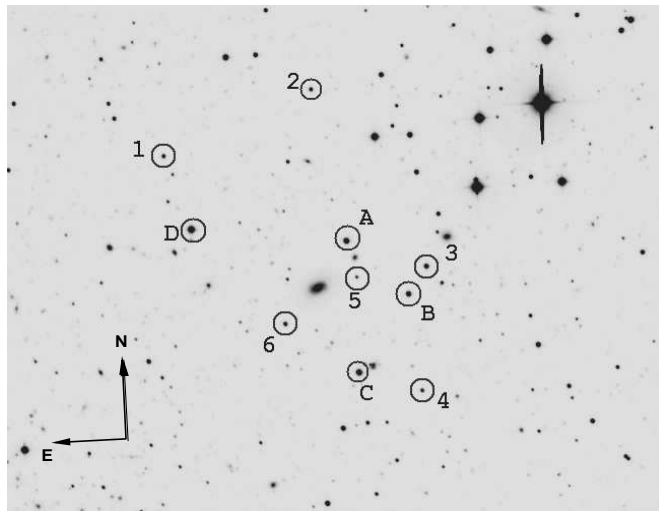}}
\caption{$I$ band field surrounding the \object{GRB 990712} host. North
  is up and east is left as indicated, and the lengths of the arrows are
  1\arcmin. The stars denoted 1--6 are secondary reference stars, and the
  stars denoted A, B, C, and D correspond to the reference stars 1, 2, 3, and
  4 in the paper of \citet{sahu00}. The GRB host itself lies very close to the
  circle surrounding star A, and is not visible in the representation here.
  The magnitudes of the stars are listed in Table~\ref{tab:712_field}.}
\label{fig:712_field}
\end{figure}

The photometric reference stars in the field were adopted from \citet{sahu00}
and no re-calibrations of the field in the $V\!RI$ bands were done. Some of
the reference stars in \citet{sahu00} were saturated in the longest DFOSC $V$,
$R$ and $I$ band exposures so the shortest exposures were analysed in order to
find secondary standard stars in the field which could be used as photometric
reference stars in the combined frame. The magnitudes were derived by
performing relative aperture photometry on the field with the PHOT package in
IRAF. A section of a DFOSC image is shown in Fig.~\ref{fig:712_field} where
the stars denoted A, B, C, and D correspond to the standard stars in
\citet{sahu00}, while the numbers represent the secondary standard stars. The
magnitudes are listed in Table~\ref{tab:712_field}.

For the $B$ band, calibration images were taken in June 2001 using the Danish
1.5 m with DFOSC. The standard field PG~1657 from \citet{landolt} was observed
4 times during the photometric night. This field contains 4 standard stars.
Only $B$ band data were taken of the field containing the host of
\object{GRB~990712}. The uncertainty of the $B$ band calibration is 8\% as
estimated by standard procedures in IRAF. The errors are propagated in
quadrature assuming that the errors are independent.

\begin{table*}
\centering  
\begin{tabular}{lllllllll}
\\ \hline \hline 
Std. &   $U$ & $B$ & $V$ & $R$ & $I$ & $J$ & $H$ & $K\!s$  \\
\hline 
1& -- &  -- & 19.84$\pm$0.02 & 18.76$\pm$0.02 & 17.48$\pm$0.03 &-- &--&--\\
2&19.16$\pm$0.07& -- & 18.21$\pm$0.01 &17.87$\pm$0.02 & 17.48$\pm$0.03&-- &--&--\\
3&--&--& 19.08$\pm$0.02     &17.94$\pm$0.02  &16.57$\pm$0.03 &-- &--&--\\
4&--&--&18.90$\pm$0.02  &18.37$\pm$0.02  & 17.86$\pm$0.03&-- &--&--\\
5&--&--&20.72$\pm$0.04  &19.73$\pm$0.02  &18.68$\pm$0.04&-- &--&--\\
6&--&--&19.21$\pm$0.01  &18.31$\pm$0.02   &17.34$\pm$0.03&-- &--&--\\
A&18.80$\pm$0.06 & 18.80$\pm$0.06 & 17.16$\pm$0.01 & 16.40$\pm$0.01& 15.64$\pm$0.01 & 14.82 $\pm$ 0.02 & 14.20 $\pm$ 0.03 & 14.04 $\pm$ 0.04\\ 
B& --& 17.91$\pm$0.05 & 16.98$\pm$0.01& 16.65$\pm$0.01 & 16.29$\pm$0.01
 & 15.91 $\pm$ 0.02 & 15.60 $\pm$ 0.04 & 15.57 $\pm$ 0.03\\
C& --& 17.69$\pm$0.07 & 16.46$\pm$0.01& 15.97$\pm$0.01 & 15.50$\pm$0.01
& 14.96 $\pm$ 0.02 & 14.51 $\pm$ 0.04 & 14.42 $\pm$ 0.04\\
D&17.69$\pm$0.07& 17.33$\pm$0.06 & 15.91$\pm$0.01 & 15.27$\pm$0.01 & 14.65$\pm$0.01  
& 13.99 $\pm$ 0.02 & 13.39 $\pm$ 0.03 & 13.28 $\pm$ 0.03\\
\hline
\end{tabular}
\caption[]{Magnitudes of the secondary standard stars. The ``--'' signs 
appear where the magnitudes for the stars have not been derived either 
because they were saturated in some images, which was the case of the 
B and C stars, or they were faint and therefore gave large photometric 
errors ($>0.08$). These were disregarded in the case of the $U$, $B$, and 
near-IR bands of the secondary standards 1--6. The $V\!RI$ band magnitudes 
of A, B, C, and D are taken from \citet{sahu00}.}
\label{tab:712_field}
\end{table*}

The $U$ band data consist of 4$\times$15 min and one 10 min exposures. The
images were reduced in the standard way and coadded using the drizzle program
without changing the pixel sizes. The night was photometric according to ESO's
monitoring of the night sky conditions at La Silla.  The zero point,
extinction coefficient in the $U$ band, and colour term of the night were
given by Ramana M. Athreya (private communication). The $U$ band magnitudes of
the reference stars A, B, C, and D were measured in the 4 individual images,
and the magnitude of the host was obtained using relative photometry in the
coadded image using a 2\farcs 4 radius aperture. This aperture was the same
used for the DFOSC images.  The resulting magnitudes of the host in all
filters are given in Table~\ref{tab:res712}. The photometric measurements were
checked with SExtractor \citep{bertin}, and with aperture photometry we found
magnitudes consistent with those in the Table within 1$\sigma$ errors.

\subsection{Ground-based near-IR images}
For the reduction of the near-IR $J\!H\!K\!s$ SOFI data, a sky image was
constructed from 6--10 object images obtained immediately before and after
each frame. The number of images depends on the quality of the sky subtraction
which was evaluated by eye. The sky subtracted images were divided by a flat
field obtained from the NTT/SOFI webpages.  A flatfield multiplied with an
illumination corrected image shows variations on the order of 0.2\% during a
month, and the reduction described here gave flatfield accuracies of 1\%.

The integration times were 31$\times$15 s, 40$\times$15 s, and 20$\times$15 s
for the $H$, $K\!s$, and $J$ band respectively. The exposures were divided
into two sets, one for each day, which were reduced separately. Standard star
observations were done right after the science exposures. The extinction
coefficients were adopted from the NTT/SOFI webpages. The corrections for
atmospheric extinction are small, because the standard star observations were
performed at an airmass different by only 0.1 from the science observations.
Likewise, the colour terms in the near-IR are small ($\lesssim$0.02). The
images of the standard star were analysed in order to find the zero point in
each filter. The transformation equations did not include a colour term, as
this produced uncertain fits, are given by: \( J_{\mathrm{inst}} = j1 +
J_{\mathrm{std}}\), where $J_{\mathrm{inst}}$ is the instrumental magnitude
and $J_{\mathrm{std}}$ is the standard magnitude. The transformations for the
$H$ and $K\!s$ data are similar.  The fitted zero points are :
$j1=23.06\pm0.005$, $h1=22.937\pm0.003$ and $k1=22.367\pm0.004$.  These zero
points agree with the values posted on the NTT/SOFI webpages.

The shifts between the reduced images were found using {\tt precor}, {\tt
  crossdrizz} and {\tt shiftfind}, and the images in each field were combined
using {\tt imcombine}. The images were combined using a common median
zero-point background value, and applying a bad-pixel mask obtained from the
SOFI webpages to reject hot or dead pixels in the combination.

The magnitudes of the reference stars (the same 4 stars as in \citet{sahu00})
were measured in 10 individual frames for each filter, and the magnitude of
the host was found by performing relative photometry to the 4 stars in the
co--added images. The magnitudes of the reference stars are presented in
Table~\ref{tab:712_field}, and the near-IR magnitudes of the host are given in
Table~\ref{tab:res712} which also presents the fluxes in the various bands.
The fluxes are obtained by correcting the magnitudes for the Galactic
reddening \ebv = 0.033 estimated from the dust maps of \citet{schlegel98}.
Then the magnitudes are converted to AB magnitudes. Due to the faintness of
the host in the near-IR, the errors are dominated by sky noise.

The near-IR magnitudes were obtained using the flux enclosed within a circular
radial aperture of 1\farcs2. At optical wavelengths, the flux of the host is
not contained within this aperture due to worse seeing. In order to get the
right colour of the host, an offset equal to the difference in the $I$ band
magnitude between an aperture of 1\farcs2 and 2\farcs4 was added. This will
then provide the right $I-J$ colour, as well as a smaller photometric error.
The same method was applied for the $H$ and $K\!s$ data. As long as the colour
gradient in the host is negligible, this does not change the near-IR
magnitudes one would have calculated from larger apertures. The colour
gradient of the host is negligible at a radius larger than 1\arcsec\,
estimated from the morphological study of the HST images described in detail
in Section~\ref{morph}. We find that there are colour gradients in the central
0\farcs25, but this has no impact on radii larger than 1\farcs2. The $K$ band
magnitude is consistent with the value reported in \citet{lefloch03}.

\begin{table}
\centering
\begin{tabular}{lll}
\\ \hline \hline 
filter  & magnitude & flux ($\mu$Jy)\\
\hline
 $U$ &  23.12$\pm$0.05 & 1.00$\pm$0.05\\
 $B$&   23.36$\pm$0.09& 2.00$\pm$0.06\\ 
 $V$&   22.39$\pm$0.03 & 4.43$\pm$0.16\\
 $R$&   21.84$\pm$0.02 & 6.31$\pm$0.20\\
 $I$&   21.41$\pm$0.03& 7.01$\pm$0.29\\
 $J$&   20.81$\pm$0.17& 7.68$\pm$1.21\\
 $H$&   20.25$\pm$0.19& 8.22$\pm$1.47\\
 $K\!s$&19.98$\pm$0.28& 6.74$\pm$1.71\\
\hline 
\end{tabular}
\caption{Magnitudes and corresponding fluxes of the host of
\object{GRB~990712} from all ground based observations. The fluxes in
column 3 are obtained by correcting for a Galactic reddening of
\ebv~=~0.033, and offsetting to the AB system before converting the
magnitudes to fluxes. The flux errors do not include the uncertainty
of the Galactic reddening. The $R$ band magnitude is consistent with that
derived in \citet{hjorth00}.}
\label{tab:res712}
\end{table}

\section{Spectroscopy}
\label{halpha}
The host galaxy of \object{GRB 990712} was observed in service mode with FORS2
at UT4 of ESO's Very Large Telescope on July 18, 2001. The exposure time was
10 min, and the grism used was GRIS 600z+23 with order separation filter
OG590, giving a wavelength range of 7400--10700~{\AA}, which includes the
redshifted wavelength of the host galaxy's H$\alpha$ at 9404~{\AA}. The slit
width was set to 1\arcsec, resulting in a resolving power of approximately
1400. The seeing during the observations was around 0\farcs8.  The spectrum
was reduced in the standard fashion, using IRAF. The wavelength calibration
was performed using a HeNeAr lamp spectrum; the resulting scatter is
0.02~{\AA}.

\begin{figure}
\centering
\resizebox{\hsize}{!}{\includegraphics[bb=50 10 670 390,clip]{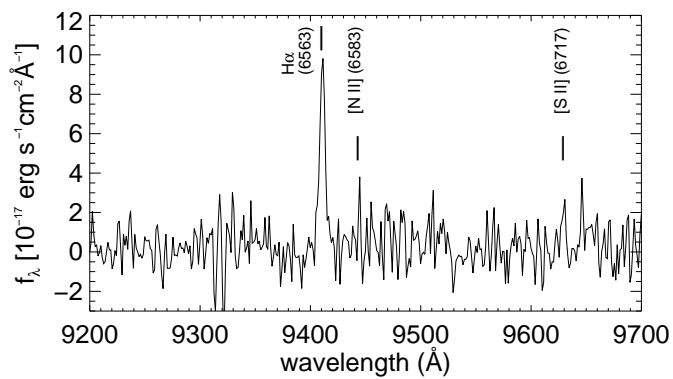}}
\caption{Section of the spectrum containing the redshifted \halpha\,
  line. The spectrum has been wavelength calibrated, and set to a flux scale
  corresponding to the continuum level of 7.38 $\mu$Jy estimated from the SED
  analysis. The redshift measured from the H$\alpha$ line is 0.4331.  At
  9436.4~{\AA} a faint emission line from [\ion{N}{II}] $\lambda$6583 is
  present, corresponding to $z=0.4338$ after applying a standard air to vacuum
  correction, and at 9621~{\AA} a line from [\ion{S}{II}] $\lambda$6717
  corresponding to $z=0.4328$ can be seen.}
\label{fig:halpha}
\end{figure}

The resulting spectrum has a bright \halpha\, emission line at 9404~{\AA}
which has an observed equivalent width (EW) of 180$\pm$40~{\AA} estimated from
fitting the continuum level by spline polynomials of different orders. A small
part of the spectrum is shown in Fig.~\ref{fig:halpha}. One clearly sees the
\halpha\, line and also at low $S/N$ levels, the [\ion{N}{II}] $\lambda$6583
line and the [\ion{S}{II}] $\lambda$6717.  The measured EW of the H$\alpha$
line indicates that the host is a young star-forming galaxy.  The starburst99
models provide a relation between the rest frame H$\alpha$ EW and the age of a
stellar population \citep{lei99}. With a rest frame EW of 125$\pm$28~{\AA} one
would expect an instantaneous starburst age of $\sim$6 Myr according to the
Starburst99 models, assuming solar metallicity.  A lower metallicity of
$Z=0.001$ would increase this age estimate by a factor of 2.  Furthermore, the
EW of H$\beta$ reported in \citet{vrees01} supports a starburst age of
$\sim$6~Myr according to the Starburst99 models.  In the case of a continuous
star formation rate of 1~M$_{\odot}$~yr$^{-1}$ the inferred age from the EW is
$\sim$60 Myr, i.e. in both scenarios the presence of a young population is
inferred.

\section{Morphology of the host galaxy}
\label{morph}

\begin{figure}
\centering
\resizebox{\hsize}{!}{\includegraphics{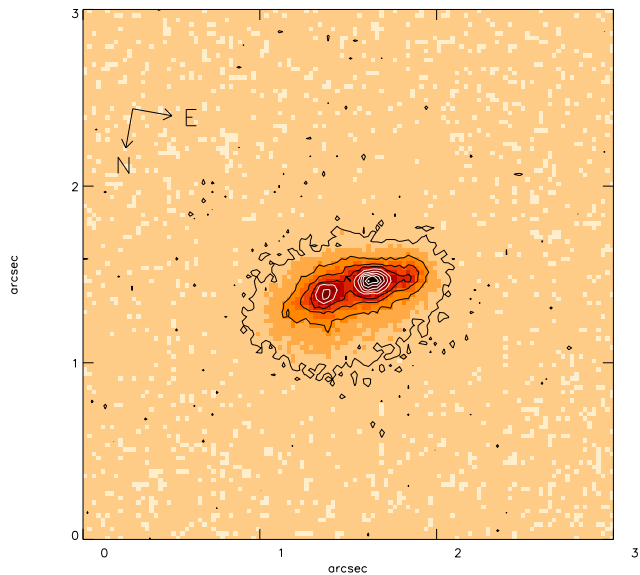}}
\caption{A STIS CL image of the host of \object{GRB 990712}
  overlayed by a contour plot shows that there are two intensity peaks in the
  image. The scale of the plot is 3\arcsec$\times$3\arcsec.  The distance
  between the two peaks is 10 drizzled pixels, or 0\farcs25, which corresponds
  to a separation of $\sim$1500~pc at $z=0.433$. At the time of observation
  more than one year from the burst the flux contribution from the afterglow
  will be negligible.}
\label{fig:712surf}
\end{figure}

In the HST images the size of the GRB host is 1\farcs3$\times$0\farcs9.  The
STIS images have revealed that the host of \object{GRB 990712} has two
separate intensity peaks, and neither of these are located at the geometrical
center of the host. An image and a contour plot of the host is shown in
Fig.~\ref{fig:712surf}. The rightmost knot (south-east=SE) is $\sim$1  mag
brighter than the left (north-west=NW). The GRB occurred in the SE knot within
0\farcs048$\pm$0\farcs080 of the center. The afterglow itself will not
contribute significantly to the flux of the host if a break in the light curve
is present around one day after the burst as suggested in \citet{gulli01}. If
no such break occurred, the contamination of the SE knot due to the late time
afterglow will be $\sim$5\%. If a supernova of similar brightness as
\object{SN~1998bw} is present at the time of the observations, it would have
the magnitude $V \approx 27.0$ at $z=0.433$. This magnitude is calculated
assuming a similar late time supernova light curve contributing to the total
flux in addition to the afterglow. If a SN is present, the contamination of
the SE knot will be an additional $\sim$5\%.

The colours of the two knots were analysed by performing aperture photometry
centered on each knot using both the CL and LP images. At a redshift of
$z=0.433$ the difference in magnitudes in these filters roughly corresponds to
the restframe $B-V$ colour. The $B-V$ colour of the NW knot is $1.3\pm0.1$,
and the colour of the SE knot is $0.9\pm0.1$ within an aperture of radius
0\farcs076 (i.e. 3 drizzled pixels). The colour becomes more red with
increasingly larger photometric apertures (up to 0\farcs15 radius, i.e. 6
drizzled pixels), but the colour gradient in the NW knot is smaller than in
the SE knot. At larger apertures, the colours will be contaminated by flux
from the other knot. The differences in colour could be caused by different
ages of two starburst regions, the bright blue knot being slightly younger
than the fainter. Another explanation could be a relatively larger extinction
in the NW knot. If the colour excess, \ebv, of the faint knot is larger than
0.34 than for the bright knot, this could explain the colour difference.
Compared to the overall extinction estimated from the SED this is relatively
large.  Even though we find no evidence for a large extinction from the SED
analysis as explained in Section~\ref{sed}, the morphology of the host
supports the presence of two different stellar components.

Most interestingly, the GRB was coincident with the center of the blue SE knot
\citep{bloom02b}. This location of the GRB corresponds to the bluest part of
the host, which likely links the GRB to a star-forming site.  If the two knots
were the result of a merging of two components, then one would expect to see
further evidence of the tidal interaction in, for example, luminous tails.
This is not apparent in the image, but for low mass systems bright tails are
probably rare. One should note that the surface brightness of tidal tails can
be low and thus difficult to detect, since the redshift of the host gives a
further factor $(1+z)^4$ dimming of the surface brightness assuming a standard
cosmology.

\section{Spectral energy distribution of the host}
\label{sed}

The magnitudes in Table~\ref{tab:res712} were used to compare to theoretical
galaxy template spectra from \citet{bru93}. This was done by the program
HyperZ\footnote{\tt http://webast.ast.obs-mip.fr/hyperz/} described in
\citet{bolzo00}. HyperZ was written mainly for obtaining the photometric
redshifts of galaxies in large surveys, but it also serves the purpose for
finding the best matching theoretical galaxy template for a given set of broad
band observations. The templates consist of elliptical, several types of
spiral, irregular, and starburst spectra at various ages having different star
formation histories. The time evolution is described by SFR $\propto$
exp(--$t/\tau$), where $\tau$ is the SFR timescale which increases along the
Hubble sequence, with $\tau\rightarrow 0$ in the case of an instantaneous
starburst. The metallicities of the templates are equal to the solar value,
$Z$=0.02. \citet{goro02,goro03} have shown that for the GRB~000210 and
GRB~000418 hosts, the metallicity is a secondary variable in comparison to the
impact of the assumed IMF and the extinction law.
  
We used both the \citet{miller79} initial mass function (IMF) and the
\citet{salpeter55} IMF for stellar masses between 0.1 and 125 M\subsun\, for
calculating the templates. The Miller \& Scalo IMF produces fewer massive
stars compared to a Salpeter IMF, and at masses below 1~M\subsun\ the Miller \&
Scalo IMF is flatter \citep{miller79}.  The largest differences between the
templates are at the red and near-IR wavelengths. Different amounts of
internal extinction can also be applied to the templates. In this way, the SED
fitting allows an estimate of the type of galaxy, age, and internal
extinction. In the analysis the redshift of the templates was fixed to the
value of the host ($z=0.433$).  Leaving the photometric redshift
$z_{\mathrm{phot}}$ as a free variable gives
$z_{\mathrm{phot}}=0.445\pm0.020$, which is consistent with the spectroscopic
one.  This additional free parameter does not change any of the resulting
values from the best fit besides changing the reduced $\chi^2$ by a small
amount.  The agreement between the photometric and the spectroscopic redshift
shows that the SED fitting technique is reliable for estimating other
properties of the host.

The goodness of the fit is evaluated by the expression:
\begin{equation}
\chi^2=\sum_i \Big( \frac{F_{\mathrm{host},i} - k \times F_{\mathrm{temp},i}}{\sigma(F_{{\mathrm{host}},i})} \Big)^2
\label{eq:chi}
\end{equation}
where the sum is to be taken over all filters, $i$. The flux values of
$F_{\mathrm{host}}$ are found in Table~\ref{tab:res712},
$\sigma(F_{{\mathrm{host}},i})$ is the photometric error, and $k$ is a
normalization constant. $F_{\mathrm{temp},i}$ is the flux of the template in
the filter $i$, which is calculated using the throughput for the given filter
and instrument. For the DFOSC and EFOSC data, the filter transmission was
convolved with the quantum efficiency of an ESO LORAL CCD in order to
calculate the throughput. For the NTT data, the combined throughput of SOFI
and the filters were used.

\begin{figure*}
\centering
\resizebox{\hsize}{!}{\includegraphics{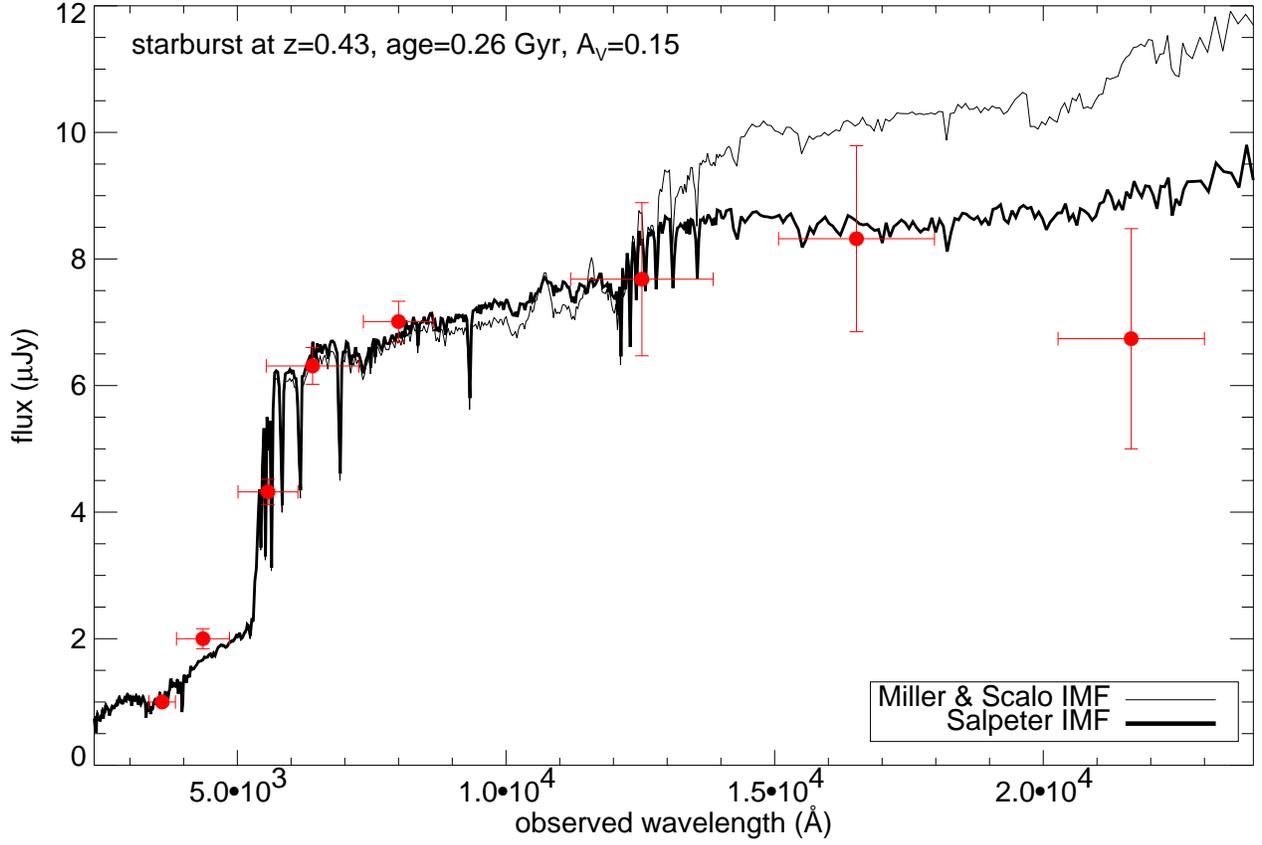}}
\caption{Best fit of the GRB host SED to synthetic spectra calculated
  by the HyperZ program. The synthetic spectra are shown by the two solid
  lines and the dots denote the $U\!BV\!RI\!J\!H\!K\!s$ fluxes from
  Table~\ref{tab:res712}. The horizontal errorbars indicate the FWHM of each
  filter. The thin line shows a template spectrum calculated using a Miller \&
  Scalo IMF while the thick line is calculated using a Salpeter IMF. The
  values of the photometric redshift, age, and extinction are given as inserts
  in the panel. A starburst extinction curve is used for the calculation of
  the extinction.}
\label{fig:hypplot}
\end{figure*}

The SED of the host was best fit by a starburst template. This confirms the
conclusion in \citet{vrees01}, derived on the basis of emission lines, that
the host is a starburst galaxy. The best fit model has a starburst age of
0.255~Gyr and an extinction of $A_V=0.15\pm 0.1$ using a Salpeter IMF for the
templates. The error in the extinction was estimated from results of the fits
for which the $\chi^2$ per degree of freedom, $\chi^2/$dof$ < 2$, and all
these fits gave an age of 0.255 Gyr.  Fig.~\ref{fig:hypplot} shows the best
fits when using templates from a Miller \& Scalo IMF (thin line) and a
Salpeter IMF (thick line) respectively.  Both templates gave the same values
for the extinction and age for the best fit.  The fit to the thin line has a
reduced $\chi^2/$dof~=~2.82 and a fit to the thick line has
$\chi^2/$dof~=~0.959. The largest difference between the two templates is in
the near-IR, where our photometric points have large uncertainties.

Fitting the SED to other types of galaxy spectra give larger values of
$\chi^2$, e.g. $\chi^2/$dof~=~14.3 for the best fit to an irregular galaxy
template, and $\chi^2$/dof~=~16.2 to an elliptical galaxy. For the latter
template, all the measured near-IR fluxes were $\sim$3$\sigma$ below the
template flux. Generally, all other templates besides starburst templates fail
to reproduce the flat continuum from 8000--22000~{\AA}, while at the same time
fitting the Balmer jump at the rest-frame 3646~{\AA}. Thus, we infer that the
host is most likely a starburst galaxy with a stellar distribution similar to
a Salpeter IMF.

The precision of the age estimation relies upon the accuracy of how well the
Balmer jump is sampled. At the redshift of the host, this jump will lie at
$\sim$5200~{\AA}. It is seen in Fig.~\ref{fig:hypplot} that with the current
set of broad band magnitudes this jump is well sampled.  Therefore, the age of
the dominant population of stars is well constrained.  However, if more than
one population of stars is present in the host, this will change the age
determination somewhat as explained in section~\ref{subsec:2pop}.

The extinction found by HyperZ is $A_V=0.15$ using the extinction curve from
\citet{calz00} appropriate for starburst galaxies. The extinction measured
from emission line widths was $A_V=3.4^{+2.4}_{-1.7}$, which is consistent
with a small extinction measured from the SED. We also tried to do the SED
fitting with other extinction curves. The $A_V$ found by HyperZ was found not
to vary much ($A_V \lesssim 0.2$ in all cases) using the extinction curve of
the Milky Way (MW) from \citet{seaton79}, the Large Magellanic Cloud (LMC)
from \citet{fitz86}, and the Small Magellanic Cloud (SMC) from
\citet{prevot84} respectively.  Acceptable values of $\chi^2/$dof$<1.5$ for
the fits were found using the starburst, LMC, and SMC extinction curves, so
the extinction curve could not be constrained from the SED. A larger
$\chi^2/$dof = 2.2 was produced using a MW extinction curve, which suggests
that the dust in the host is different than from MW dust. This is in agreement
with results obtained from other GRBs, where the extinction law has been
inferred from studying the afterglows.  \citet{jensen01}, \citet{fynbo01b},
\citet{lee01}, and \citet{holland02b} find that the SMC extinction law gives a
better fit to multiband observations of the afterglows of
\object{GRB~000301C}, \object{GRB 000926}, \object{GRB 010222}, and
\object{GRB~021004}, respectively.

From the SED of the host we can determine the absolute magnitude in various
bands by convolving the restframe spectrum of the best fitting template with
various filter transmission curves. This gives $M_B$~=~$-$18.8, $M_V=-19.9$,
and $M_R$~=~$-$20.3 in the assumed cosmology, corrected for Galactic
extinction.  This absolute magnitude is comparable to that of other hosts
\citep{bloom98b,bloom01,sok01}.  The luminosity of the galaxy is less than the
characteristic luminosity $L^*$ given by the Schechter function. For field
galaxies a value of $M_B^*~=~-21$ is typically assumed, while for starburst
galaxies at high redshift this magnitude is highly uncertain, but is likely to
be brighter than --21 \citep{lilly95}. GRB hosts have been shown typically to
be under-luminous \citep{lefloch03}.  From the early optical light curves of
the optical transient, obtained in the $V$, $R$, and $I$ bands when it was
still bright, \citet{sahu00} estimated the luminosity of the underlying host
to be of the order of $L^*$. Using high spatial resolution HST images
\cite{hjorth00} refined the luminosity of the galaxy to be 0.2$L^*$.

\subsection{Two population fits}
\label{subsec:2pop}
The scenario described above, where the whole spectrum of the galaxy can be
represented by one single burst of star formation is likely too simple. One
must expect that more than one burst of star formation would contribute to the
total mix of stars observed. With the knowledge that some GRBs originate from
collapsing massive stars, we would expect a population of younger stars to be
present, which is supported by the large EW of the H$\alpha$ line. We
therefore investigated whether the SED of the host could be explained by a
superposition of two populations of stars, i.e. a young burst superimposed
onto an older population.

The method applied was as follows. We created two new SEDs, a red and a blue
one, whose sum was the total SED of the observed galaxy. The two objects were
run though HyperZ, finding the best fit templates using the same templates as
in the single population case. The best fit spectra of the two populations
were summed and compared to the broad band fluxes of the host, and the
$\chi^2$s of the fits were calculated using Eq.~(\ref{eq:chi}). It was then
investigated by iterations if a bluer first population plus a redder second
population would produce a better fit to the observations.

This process was done first with two populations of similar total flux. In a
second run, the observed flux was partitioned into 80\% for the first
population and 20\% for the second.

\begin{figure}
  \centering \resizebox{\hsize}{!}{\includegraphics[bb=50 10 650 390,clip]{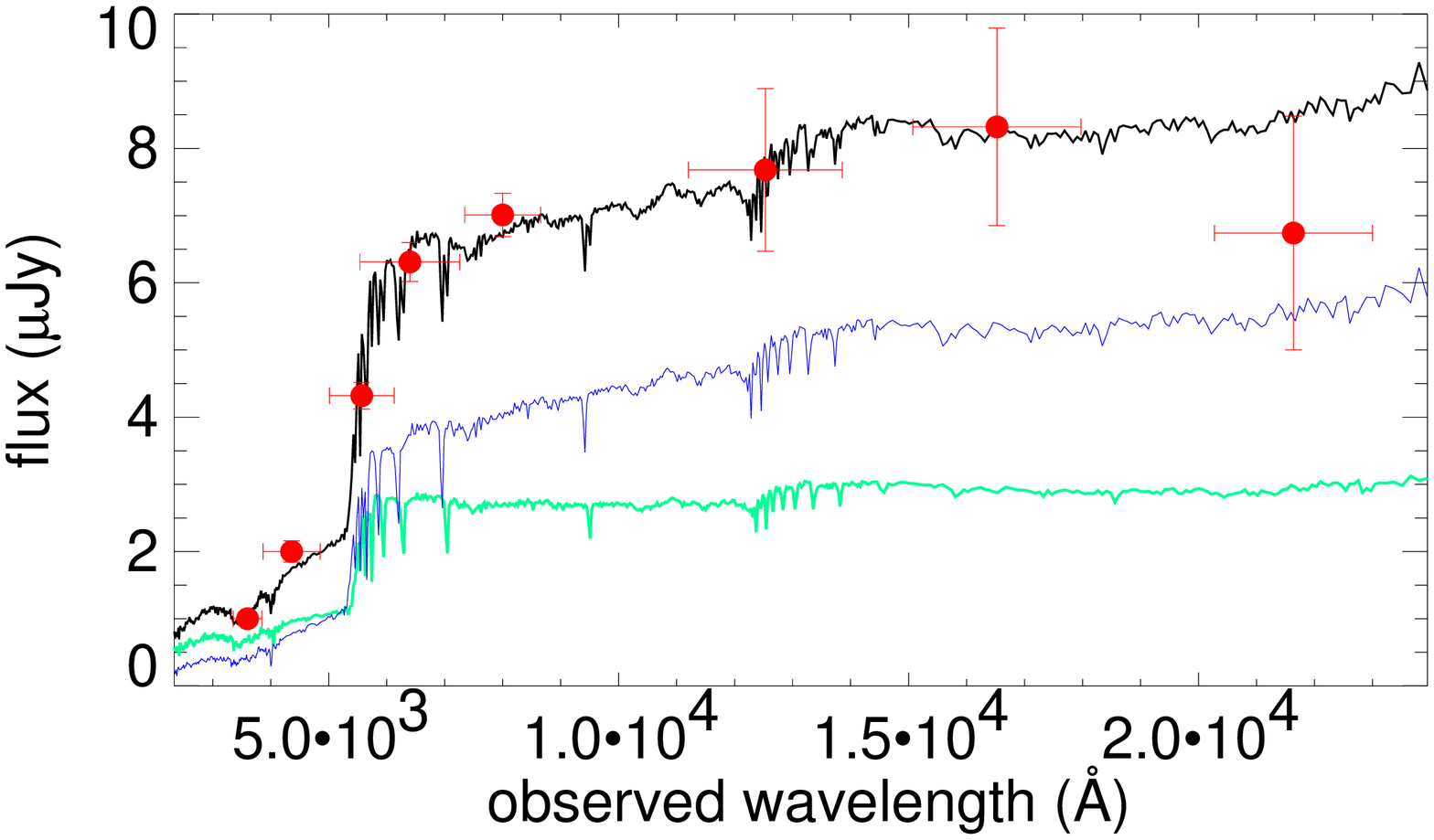}}
  \resizebox{\hsize}{!}{\includegraphics[bb=50 10 650 390,clip]{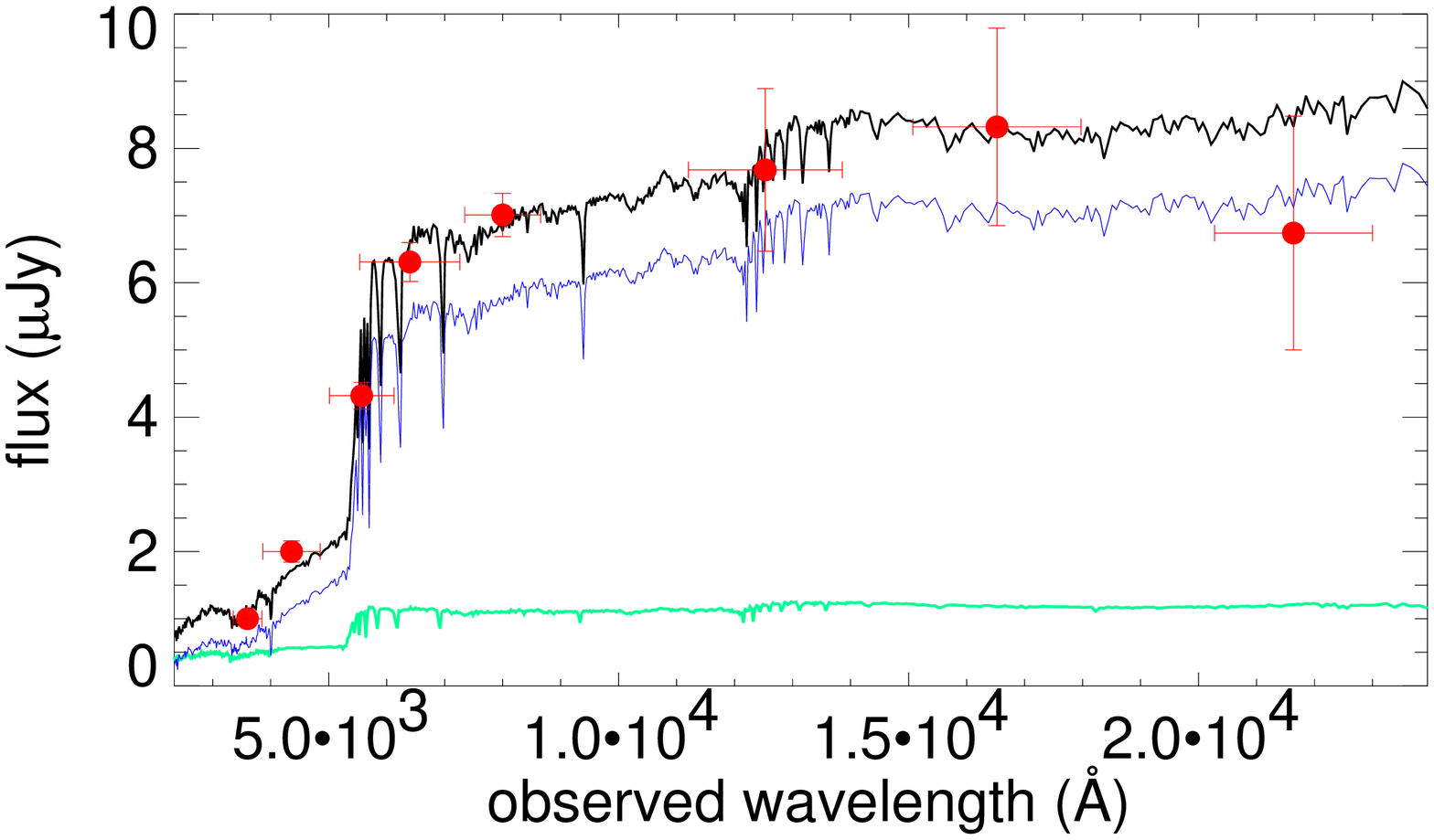}}
\caption{Two examples of the best fits for two populations. In both
  panels the best fit spectra for each of the components are shown along with
  the summed spectrum. In the upper panel, the lower spectrum corresponds to a
  0.18 Gyr starburst population, the middle spectrum to a 0.36 Gyr starburst
  population, and the upper spectrum to a sum of the two. The fit has
  $\chi^2$/dof~=~0.66. In the lower panel, the lower spectrum corresponds to a
  52 Myr starburst population, the middle spectrum to a 0.36 Gyr starburst
  population, and again the upper spectrum to a sum of the two. This fit has
  $\chi^2$/dof~=~0.75.}
\label{fig:2pop1}
\end{figure}

We found that several two population models were able to fit the observations
with $\chi^2$/dof~$\approx$~1. As shown in the upper panel in
Fig.~\ref{fig:2pop1}, the total flux can originate from two rather similar
instantaneous burst populations. In this specific case, one of the populations
has an age of 0.36 Gyr and an extinction of $A_V=0.12$, while the other has an
age of 0.18 Gyr and zero extinction. The fit of the summed spectrum to the
observations is $\chi^2$/dof~=~0.66.
 
Another scenario could also explain the properties of the host. The lower
panel in Fig.~\ref{fig:2pop1} shows the result of the first population of
stars being a 0.36 Gyr old starburst, while the flatter spectrum corresponds
to a less luminous second population, which is a 52 Myr starburst. Both these
spectral templates have $A_V$~=~0.00. The fit of the summed spectra to the
observations gives $\chi^2$/dof~=~0.75.  We therefore conclude that if a second
population of stars has a significant contribution to the total flux, the age
will be 50--200 Myr, found from acceptable values of the $\chi^2$/dof fits.
Considering the large observed H$\alpha$ EW we find that a young population
($\sim$50 Myr) is preferable.  Even younger ages for the second population can
not be ruled out. By constructing templates of very young populations of
stars, we can estimate the total flux allowed from such a population. The
constraining factor is the weak blue continuum observed. If a stellar
population with an age of 10Myr is present, the total (bolometric) flux it
emits is less than 5\% of the total observed flux from the host.

\section{Star-formation rate}
\label{SFR}

The continuum in the UV part of the spectrum (1500--2800~{\AA}) mainly comes
from young OB stars, and a relation between the UV flux and the SFR can be
derived by comparison of observed spectra to synthetic model spectra (K98).

We interpolate between the observed flux in the $U$ and the $B$ band, and the
flux at the wavelength 2800$(1+z)$~{\AA} is estimated assuming a powerlaw
spectrum, $F_{\mathrm{host}}\propto \nu^{\beta}$.  We do not consider any best
template fit for the calculation of the flux at this wavelength.  The total
luminosity at $2800(1+z)$~{\AA} can be calculated given the cosmological
model, and converted to an overall SFR of the host. K98 gives the relation
between the UV luminosity and the SFR for a Salpeter IMF.

\begin{equation}
\textrm{SFR\,(M}_{\odot} \peryr) = 1.4 \times 10^{-28} \times
L_{\nu,\mathrm{UV}} \quad [\textrm{erg}\ \  \textrm{s}^{-1}\ \textrm{Hz}^{-1}]
\label{eq:sfruv}
\end{equation}

This expression is adequate for galaxies with constant stellar formation over
time scales of at least 10$^8$ years. The coefficient that links the SFR with
the UV luminosity (1.4$\times$10$^{-28}$ in the case of Eq.~(\ref{eq:sfruv}))
is significantly lower in younger stellar populations present in starburst
regions.  Thus, this SFR has to be considered as an upper limit.  However, in
starburst regions dust is expected to be present, and a correction for dust
extinction is necessary in order to find the UV flux. We have in the SED
analysis found only a small global extinction effect, but it could be locally
higher if the medium is clumpy.  For instance, the appreciable extinction
inferred along the line of sight to some afterglows derived from the curvature
of the near-IR/optical afterglow SEDs in e.g.  GRB 000301C in
\citet{jensen01}, and GRB 000926 in \citet{fynbo01b} seem to indicate that GRB
progenitors tend to be embedded in dusty local regions.  High $N_H$ values
derived from X-ray afterglow spectra also supports this scenario
\citep{piro01}.  This will lead to an underestimate of the local extinction,
and hence of the SFR.  This opposite effect may mitigate the SFR
overestimation entailed in Eq.~(\ref{eq:sfruv}). The estimated SFR value has
to be considered as the sum of the SFR contributions coming from the two
potential stellar populations.

We calculate a flux at the rest frame 2800~{\AA} of 1.65$\pm$0.04~$\mu$Jy which
translates into a SFR of 1.3$\pm$0.3~M$_{\odot}$~\peryr, which is similar to
the SFR of a galaxy such as the Milky Way. The errors due to the flux
measurements and interpolation between the two bands are insignificant
compared to the uncertainty of the conversion factor, which is $\sim$30\%.
Correcting for the extinction of $A_V=0.15$ using the starburst extinction
curve from \citet{calz00} gives a slightly larger flux at 2800 {\AA} in the
rest frame, and the SFR is a bit larger: 1.6$\pm$0.3~M$_{\odot}$ \peryr.

We also estimate the SFR from the \halpha\ line flux. K98 gives the relation
between the \halpha\, luminosity and the SFR:
\begin{equation}
\textrm{SFR\,(M}_{\odot} \peryr) = 7.9 \times 10^{-42} \times L_{\halpha}\quad
[\textrm{erg} \ \ \textrm{s}^{-1}]
\end{equation}
We can find the line flux, $f_{\mathrm{line}}$, and the luminosity,
$l_{\mathrm{line}}$, once the continuum flux $f_{\mathrm{cont}}$ at the
wavelength of the redshifted \halpha\, line $\lambda$, and the equivalent
width $EW$ is known.
 
\begin{eqnarray}
f_{\mathrm{line}}&=& f_{\mathrm{cont}} \times c/\lambda^2 \times 10^{-10} \times EW \nonumber \\
l_{\mathrm{line}}&=& f_{\mathrm{line}} 4\pi d_L^2
\label{eq:fline}
\end{eqnarray}

Since the spectrum has not been flux calibrated, we estimate the continuum
level by interpolation between the $I$ and $J$ band fluxes assuming a powerlaw
spectrum. We find a continuum level of $(7.38\pm0.44)~\mu$Jy. Using
Eq.~(\ref{eq:fline}) we find
$f_{\mathrm{line}}=(4.50\pm1.00)\times10^{-16}$~erg~s$^{-1}$ cm$^{-2}$, which
gives a SFR=(2.8$\pm$0.7)~M$_{\odot}$~\peryr. The error includes the
uncertainty of the conversion from flux to SFR, which is 30\%. The two
calculated SFRs are consistent with each other within 2$\sigma$.

The larger SFR inferred from the \halpha\, line compared to the SFR found from
the UV flux could be due to dust extinction, which is stronger in the UV
region.  The ratio between the \halpha\, line flux and the \hbeta\, line flux,
(1.33$\pm$0.20)~$\times10^{-16}$~erg~cm$^{-2}$~s$^{-1}$ in \citet{vrees01}, is
3.38$\pm$0.90, while the expected ratio in \ion{H}{II} regions in the case of
no extinction is 2.85 \citep{osterbrock89}. The ratio between the observed
line fluxes thus corresponds to a magnitude difference of $0.19\pm0.31$. Using
the extinction curve from \citet{calz00} an extinction of $A_V~=~0.60\pm0.99$
is inferred from the line ratio. The calculated extinction is the same when
using the MW extinction curve from \cite{fitz99}, while in the case of an SMC
extinction curve one would find $A_V~=~0.41\pm0.60$.  The extinction is
therefore consistent with the small value indicated by the SED analysis.

The SFR of this host is in the same range as the SFRs found for other hosts
through their rest frame UV flux, which typically gives SFRs $<10~$M\subsun
\peryr\, \citep{fruchter99,bloom98b,djor01b}. The largest SFRs found from
optical methods to date are 20~M\subsun\,\peryr\, for the \object{GRB~990703}
host \citep{djor98} and 55~M\subsun\,\peryr\ for the \object{GRB~000418} host
\citep{bloom02c}, although a smaller SFR based on the UV region of the latter
host has been inferred \citep{goro03}. It must be pointed out that these SFRs
are strictly lower limits to the true SFRs since the reported values are not
corrected for extinction by dust in the hosts.  Radio and sub-mm data suggest
SFRs one or two orders of magnitude larger than the optical inferred SFRs for
a sample of GRB hosts \citep{berger02}.

\section{Discussion and conclusions}
\label{disc}
From broad band magnitudes in $U\!BV\!RI\!J\!H\!K\!s$ filters we have examined
the SED of the host of \object{GRB 990712}. Comparing this SED with model
templates of different galaxy types, we found that the host is a starburst
galaxy with an extinction of $A_V~=~0.15$.  With spectroscopic observation of
the host we calculated the extinction $A_V=0.6\pm 0.99$ from the
\halpha/\hbeta\, line ratio, confirming a small extinction value.

In the collapsar scenario the progenitor of the GRB may be embedded in a
molecular cloud having a much larger extinction due to the surrounding dust.
Thus, even though we can estimate the overall internal extinction in the host
in the particular case of GRB 990712, it is not possible to say anything about
the extinction in the line of sight towards the burst itself. It could well be
much higher. However, analyses of several afterglows have failed to reveal a
very high extinction ($A_V~>~1$). According to \citet{gal01}, the expected
visual extinction of the afterglows should be much higher when compared to the
column density inferred from the X-ray afterglows of several bursts. They
argued that dust can be destroyed along the line of sight towards the burst
making the visual extinction appear smaller.  \citet{wax00} have calculated
that dust can be destroyed out to a distance of 10 pc from the burst site. The
extinction inferred from the SED fitting is an overall average extinction of
the entire galaxy. Therefore it is necessary to investigate further the
relation between the small extinction inferred from the optical light curves
of some GRBs, $A_V<0.2$ \citep{andersen00,gal01,jensen01,fynbo01b,stanek01},
and the extinction in the host itself.

The SED of the host is similar to that of a starburst population with an age
of 0.26 Gyr at a redshift of 0.43. This age is still consistent with a merging
neutron star scenario as the progenitor of the GRB. It is now known that some
of the long-duration GRBs are associated with collapsing massive stars
\citep{stanek03,hjorth03}. Considering that the life times of the most massive
stars are of the order of a few Myr, a small age of the star burst is
expected.  It was therefore investigated whether two distinct populations were
able to fit the broad band observations of the host. It was found that in such
case, the best fits were produced by a younger starburst population with an
age of 50--180 Myr, and zero extinction. We consider the lower limit to be
more likely given the large H$\alpha$ EW found in the spectrum, which suggests
presence of a young stellar population with an age of 6--60 Myr depending on
the star formation history. 

From the analysis of the SED we found that a Salpeter IMF was able to
reproduce galaxy spectral templates corresponding to the observed fluxes.

We calculate the SFR by estimating the rest frame flux at 2800~{\AA}. The
SFR=1.3$\pm$0.30~M$_{\odot}$~\peryr\, is not very large, and correcting for
internal extinction in the host does not increase the SFR much.  This
relatively small SFR is comparable to that of other GRB hosts found from using
the same UV--SFR estimator. Considering that the host is less luminous than an
$M^*$ galaxy, this SFR is relatively high compared to present day galaxies.
As the host is a 0.2$L^*$ galaxy we find a SFR per $L/L^*$ of
$\sim$5~M$_{\odot}$\peryr\ $(L/L^*)^{-1}$.

Comparison with the SFR of 2.8$\pm$0.7~M$_{\odot}$~\peryr\, found from the
\halpha\, line flux implies that there may be moderate extinction present in
the host. The dust may be distributed in a clumpy medium, where most of the UV
flux is absorbed.  

The analysis of the morphology of the host showed that it has two knots of
different colours. This colour difference could be due to two bursts of star
formation. This interpretation is consistent with the large Balmer break in
the SED which suggests the presence of an older population together with the
large \halpha\ EW suggesting the presence of a very young population of stars.
  
Most importantly, the location of the burst was in the bluest part of the host
galaxy, which supports the recent observations that the long-duration GRBs are
linked to sites of formation of massive stars.

\begin{acknowledgements}
  The observations from the Danish 1.5m Telescope were supported by the Danish
  Natural Science Research Council through its Center for Ground Based
  Observational Astronomy (IJAF). This work was supported by the Danish
  Natural Science Research Council (SNF). J. Gorosabel acknowledges the
  receipt of a Marie Curie Research Grant from the European Commission. We are
  grateful for the availability of the WCS pipelines provided by Andreas
  Jaunsen. Many thanks to Jeremy Bailin (Steward Observatory, University of
  Arizona) for useful comments on the paper.

\end{acknowledgements}

\bibliography{H4202}

\end{document}